\documentclass[aps, prl, twocolumn, superscriptaddress, longbibliography, floatfix]{revtex4-1}
\usepackage{natbib}
\usepackage{amsmath} 
\usepackage{amssymb}
\usepackage[pdftex]{graphics}
\usepackage{graphicx}
\usepackage{times}
\usepackage{wasysym}

\newcommand{\ceir}{Ba$_3$CeIr$_2$O$_9$}
\newcommand{\tiirthree}{Ba$_3$Ti$_{2.7}$Ir$_{0.3}$O$_9$}

\begin{document} 

\title{Resonant inelastic x-ray incarnation of Young's double-slit experiment}

\author{A. Revelli}
\affiliation{II. Physikalisches Institut, Universit\"{a}t zu K\"{o}ln, Z\"{u}lpicher Strasse 77, D-50937 K\"{o}ln, Germany}
\author{M. Moretti Sala}
\affiliation{European Synchrotron Radiation Facility, BP 220, F-38043 Grenoble Cedex, France}
\affiliation{Dipartimento di Fisica, Politecnico di Milano, Piazza Leonardo da Vinci 32, I-20133 Milano, Italy}
\author{G. Monaco}
\affiliation{Dipartimento di Fisica, Universit\`{a} di Trento, via Sommarive 14, 38123 Povo (TN), Italy}
\author{P. Becker}
\author{L. Bohat\'{y}}
\affiliation{Abteilung Kristallographie, Institut f\"{u}r Geologie und Mineralogie, Z\"{u}lpicher Strasse 49b, D-50674 K\"{o}ln, Germany}
\author{M. Hermanns}
\affiliation{Institut f{\"u}r Theoretische Physik, Universit\"{a}t zu K\"{o}ln, Z\"{u}lpicher Strasse 77, D-50937 K\"{o}ln, Germany}
\affiliation{Department of Physics, University of Gothenburg, SE-412 96 Gothenburg, Sweden}
\affiliation{Department of Physics, Stockholm University, AlbaNova University Center, SE-106 91 Stockholm, Sweden}
\affiliation{Nordita, KTH Royal Institute of Technology and Stockholm University, Roslagstullsbacken 23, SE-106 91 Stockholm, Sweden}
\author{T.C.~Koethe}
\author{T. Fr\"{o}hlich}
\author{P. Warzanowski}
\author{T. Lorenz}
\affiliation{II. Physikalisches Institut, Universit\"{a}t zu K\"{o}ln, Z\"{u}lpicher Strasse 77, D-50937 K\"{o}ln, Germany}
\author{S.V.~Streltsov}
\affiliation{M.N.\ Miheev Institute of Metal Physics, Ural Branch, Russian Academy of Sciences, 620137 Ekaterinburg, Russia}
\affiliation{Ural Federal University, Mira Street 19, 620002 Ekaterinburg, Russia}
\author{P.H.M. van Loosdrecht}
\author{D.I. Khomskii}
\affiliation{II. Physikalisches Institut, Universit\"{a}t zu K\"{o}ln, Z\"{u}lpicher Strasse 77, D-50937 K\"{o}ln, Germany}
\author{J. van den Brink}
\affiliation{Institute for Theoretical Solid State Physics, IFW Dresden, Helmholtzstrasse 20, 01069 Dresden, Germany}
\author{M. Gr\"{u}ninger}
\affiliation{II. Physikalisches Institut, Universit\"{a}t zu K\"{o}ln, Z\"{u}lpicher Strasse 77, D-50937 K\"{o}ln, Germany}

\begin{abstract}
	Young's archetypal double-slit experiment forms the basis for modern diffraction techniques: 
	the elastic scattering of waves yields an interference pattern that captures the real-space structure. 
	Here, we report on an inelastic incarnation of Young's experiment and demonstrate that 
	{\it resonant inelastic} x-ray scattering (RIXS) measures interference patterns 
	which reveal the symmetry and character of electronic excited states in the same way 
	as elastic scattering does for the ground state. 
	A prototypical example is provided by the quasi-molecular electronic structure of insulating Ba$_3$CeIr$_2$O$_9$ 
	with structural Ir dimers and strong spin-orbit coupling. 
	The double 'slits' in this resonant experiment are the highly localized core levels of 
	the two Ir atoms within a dimer. 
	The clear double-slit-type sinusoidal interference patterns that we observe allow us to characterize 
	the electronic excitations, demonstrating the power of RIXS {\it interferometry}
	to unravel the electronic structure of solids containing, e.g., dimers, trimers, ladders, 
	or other superstructures.
\end{abstract}

\date{January 18, 2019}

\maketitle

\section{Introduction} 

RIXS provides a powerful example of particle-wave duality in quantum mechanics. 
In RIXS in the particle picture, 
an incident x-ray photon excites an electron out of the core of an atom into an empty valence level. 
The highly excited atomic state that is produced in this way contains an extremely localized hole 
in its core, with a size of a few pm. 
Subsequently this intermediate state decays: a valence electron fills 
the core hole under reemission of a photon with lower photon energy than the incident one.
The final excited state may correspond to, e.g., an interband, orbital, or magnetic excitation \cite{Ament11}.
Here, we focus on the equivalent picture of x-ray waves that are scattered via a localized intermediate 
state and interfere.

In the early 1990s it was realized that even if in RIXS the scattering is inelastic and the atomic core hole 
is very local, the amplitudes for its creation and annihilation have to be summed up {\it coherently} 
when identical ions are involved over which the {\it final} excited state is delocalized \cite{Ma94,Ma95,Gelmukhanov94}.
As a consequence, interference effects become possible.
Specifically, in 1994 an interference pattern equivalent to Young's double-slit experiment was 
predicted for RIXS on diatomic molecules \cite{Ma95,Gelmukhanov94}, 
an effect that so far has not been observed.  
A double-slit-like interference occurs because the RIXS intermediate state contains a single core hole 
that can be on either of the two atoms in the molecule, see Fig.~1. 
The final state exhibits an electron in an excited molecular orbital which is delocalized over the two atoms. 
It corresponds to an observer without `which-path' information, i.e., one cannot tell on which atom 
the core hole was localized in the intermediate state. 
The emitted x-rays interfere and give rise to a double-slit-type sinusoidal interference pattern as a function 
of the momentum {\bf q} that is transferred in the inelastic scattering process.

\begin{figure}[b]
	\centering
	\includegraphics[width=0.65\columnwidth]{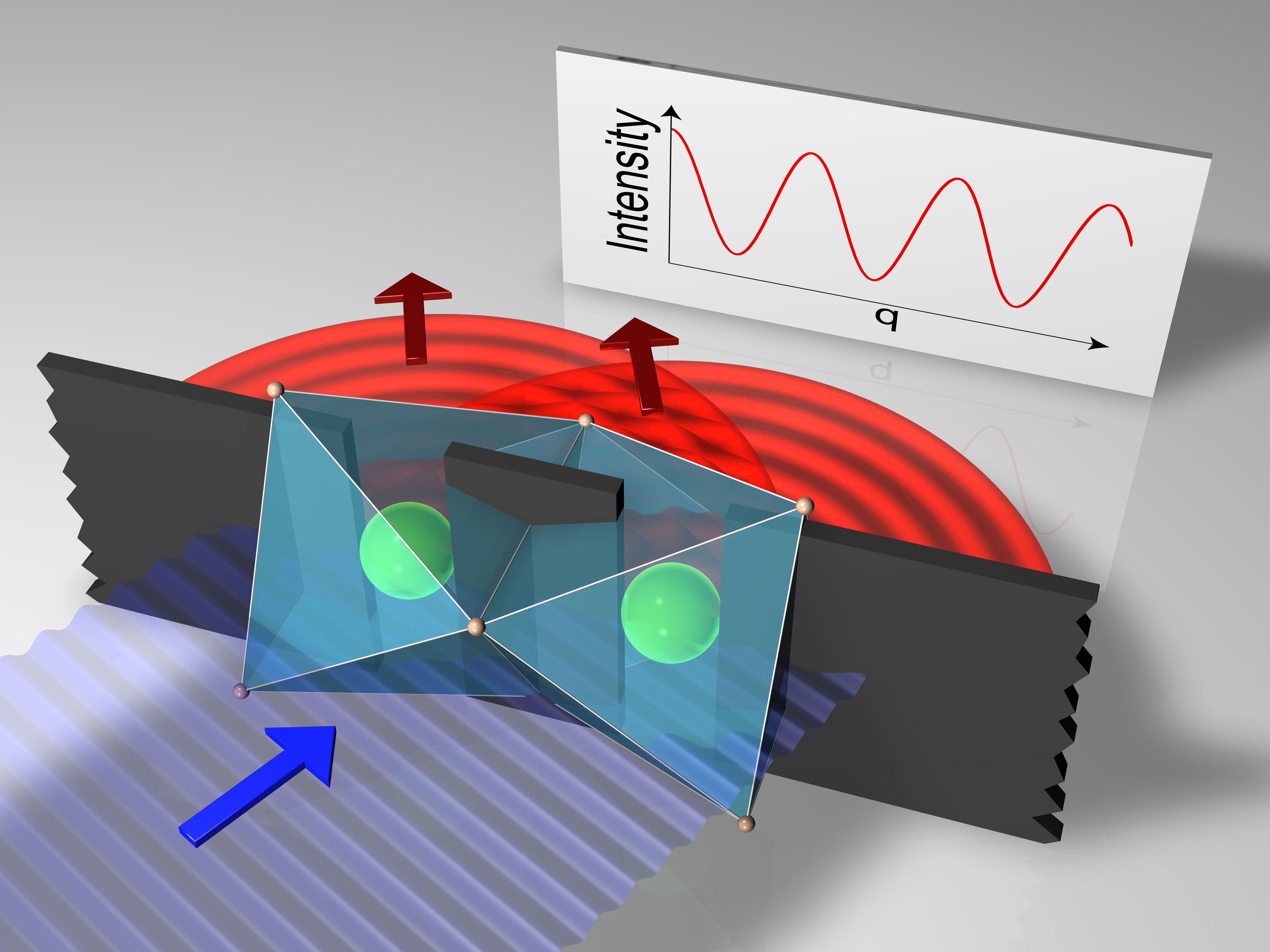}
	\caption{\textbf{Double-slit-type RIXS  interferometry for an Ir$_2$O$_9$ bioctahedron.}
		The incident plane wave (blue) resonantly excites a core electron on one of two equivalent Ir sites at 
		$\mathbf{r}_1$ and $\mathbf{r}_2$.	This intermediate state decays to a quasi-molecular final state which 
		is delocalized over both sites, i.e., without `which-path' information. 
		The emitted x-rays interfere with each other, giving rise to a double-slit-type sinusoidal interference pattern 
		as a function of the transferred momentum $\mathbf{q}$ which points along $\mathbf{r}_1-\mathbf{r_2}$.
	}
	\label{fig_double}
\end{figure}

In a gas of diatomic molecules, the interference is blurred by orientational disorder. 
This complication is absent in crystalline solids with a quasi-molecular electronic structure such as 
insulating Ba$_3$CeIr$_2$O$_9$ (BCIO, see Fig.~2A) \cite{Doi04}, an ideal model system 
with quasi-molecular orbitals localized on well-ordered structural dimers. As we show below, 
the excellent energy and momentum resolution of state-of-the-art RIXS \cite{Moretti18} 
allows us to observe astonishingly clear interference patterns which are the unambiguous fingerprints 
of the symmetry of the low-energy electronic excitations.

\section{Results}

We have grown single crystals of hexagonal BCIO (space group $P6_3/mmc$) by the melt-solution technique, 
see \textit{Methods}. 
Each of the Ir$^{4+}$ ions within the structural dimers formally shows a $5d^5$ configuration, 
with one hole in the $t_{2g}$ shell. However, the nearest-neighbor Ir-Ir distance of 2.5 \AA{} is even shorter 
than the 2.7 \AA{} found in Ir metal. Accordingly, the intra-dimer Ir-Ir hopping is large, 
driving the formation of quasi-molecular orbitals with large bonding-antibonding splitting. 
It should be stressed that this situation is very different from the case of a single Ir$^{4+}$ site, 
where strong spin-orbit coupling ($\lambda \approx 0.4 - 0.5$ eV) splits up the local $t_{2g}$ manifold 
and yields spin-orbit-entangled $j=1/2$ moments, see Fig.~2B. 
Prominent examples showing rich $j=1/2$ physics are Sr$_2$IrO$_4$ \cite{Kim08,Kim09,RauReview,Kim12}
and Na$_2$IrO$_3$ \cite{Chaloupka10,Katukuri14,Gretarsson13PRL}.

\begin{figure}[b]
	\centering
	\includegraphics[width=1\columnwidth]{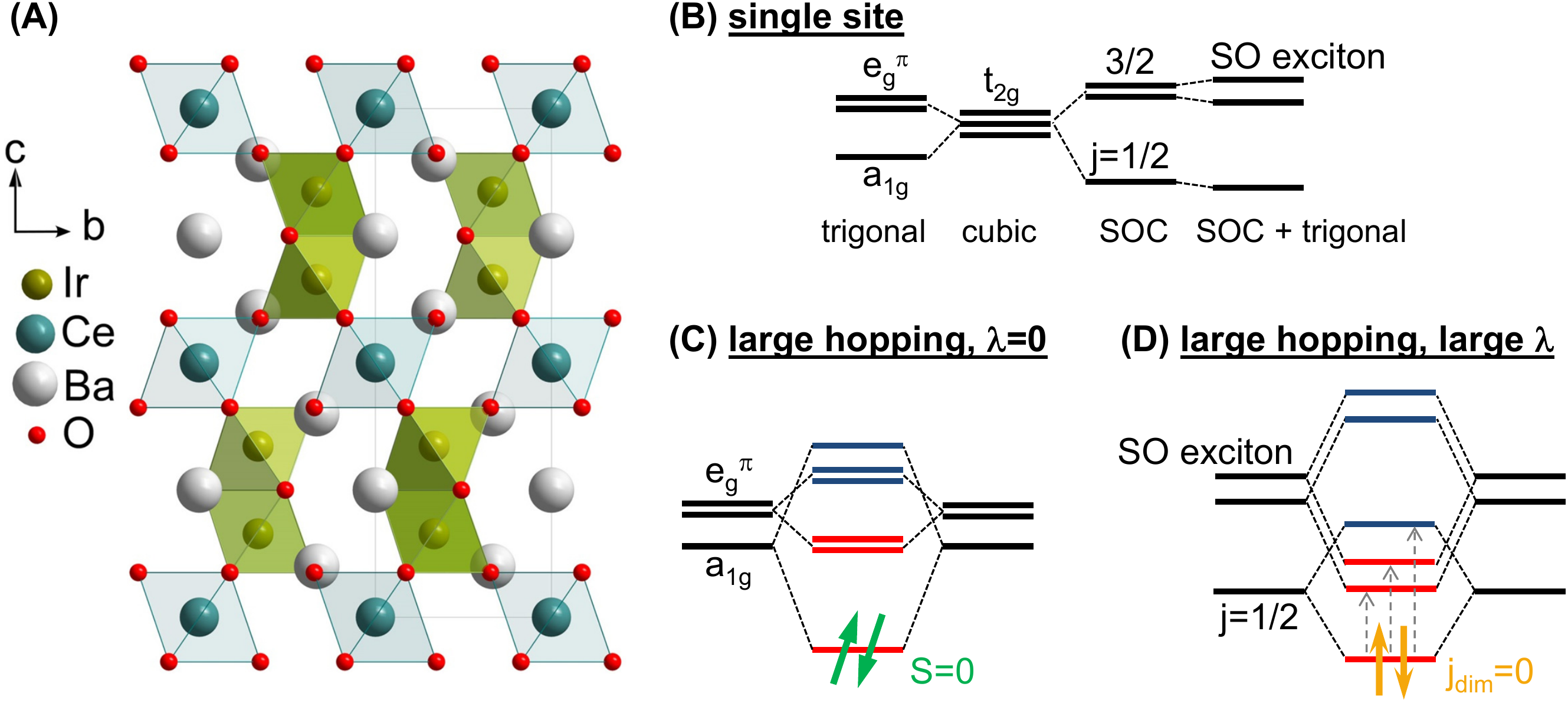}
	\caption{\textbf{Crystal structure and quasi-molecular orbitals of Ba$_3$CeIr$_2$O$_9$.} 
		(\textbf{A}) Layers of Ir$_2$O$_9$ bioctahedra (light green) are sandwiched between Ce layers. 
		The two Ir ions are displaced by 2.5~\AA{} along $c$. 
		(\textbf{B})
		All sketches refer to the hole representation. 
		For a single Ir$^{4+}$ site, the $t_{2g}$ level is split by a trigonal crystal field
		$\Delta_{\rm CF}$ into $a_{1g}$ and $e_g^\pi$ orbitals 
		or by spin-orbit coupling $\lambda$ (SOC) into $j$\,=\,$1/2$ and $3/2$ states.
		Real materials show both, $\lambda$ and $\Delta_{\rm CF}$, which yields three distinct orbitals (right), 
		where the two excited states are called spin-orbit exciton.
		(\textbf{C,D}) Sketches of quasi-molecular
		orbitals for two Ir$^{4+}$ sites with dominant hopping and small $\Delta_{\rm CF}$. 
		Bonding and antibonding levels \cite{Streltsov16,KhomskiiZhETF} are depicted in red and blue, respectively. 
		In (\textbf{C}), $\lambda$\,=\,$0$, as on the left hand side in (\textbf{B}). 
		The ground state is the $a_{1g}^2$ spin singlet with total $S$\,=\,$0$ (green arrows). 
		In (\textbf{D}), both spin-orbit coupling and hopping are large, as on the right hand side in (\textbf{B}).
		(Anti-)Bonding states are formed from local $j$ states.
		The ground state is a total $j_{\rm dim}=0$ singlet built from two $j$\,=\,$1/2$ states (orange arrows). 
		Vertical dashed arrows indicate the 3 lowest excitations, which correspond 
		to peaks $\alpha$, $\beta$, and $\gamma$ in the RIXS spectra. 
		Rigorous calculations of the eigenstates (see methods and appendix) support that the simple picture 
		plotted in (\textbf{D}) contains the essential character of the low-energy excitations. 
	}
	\label{fig_struc}
\end{figure}

There are two limiting scenarios for an effective description of the electronic structure of BCIO, i.e., 
for the character of the bonding and antibonding quasi-molecular orbitals. 
For strong spin-orbit coupling, (anti-) bonding states can be formed from spin-orbital $j=1/2$ states, see Fig.~2D. 
However, a large Ir-Ir hopping may quench the $j=1/2$ moments, as has been discussed for Na$_2$IrO$_3$ 
\cite{Mazin12,Foy13}. In this case, the crystal-field-split $t_{2g}$ orbitals provide a more appropriate basis 
for the formation of (anti-) bonding states, see Fig.~2C. 
As will become clear in the following, 
these substantial differences in the Ir $5d$ orbitals can be highlighted and quantified 
by RIXS  interferometry, i.e., RIXS measurements of {\bf q}-dependent interference patterns 
which reveal the symmetry and character of the excited states.

\begin{figure}[b]
	\centering
	\includegraphics[width=\columnwidth]{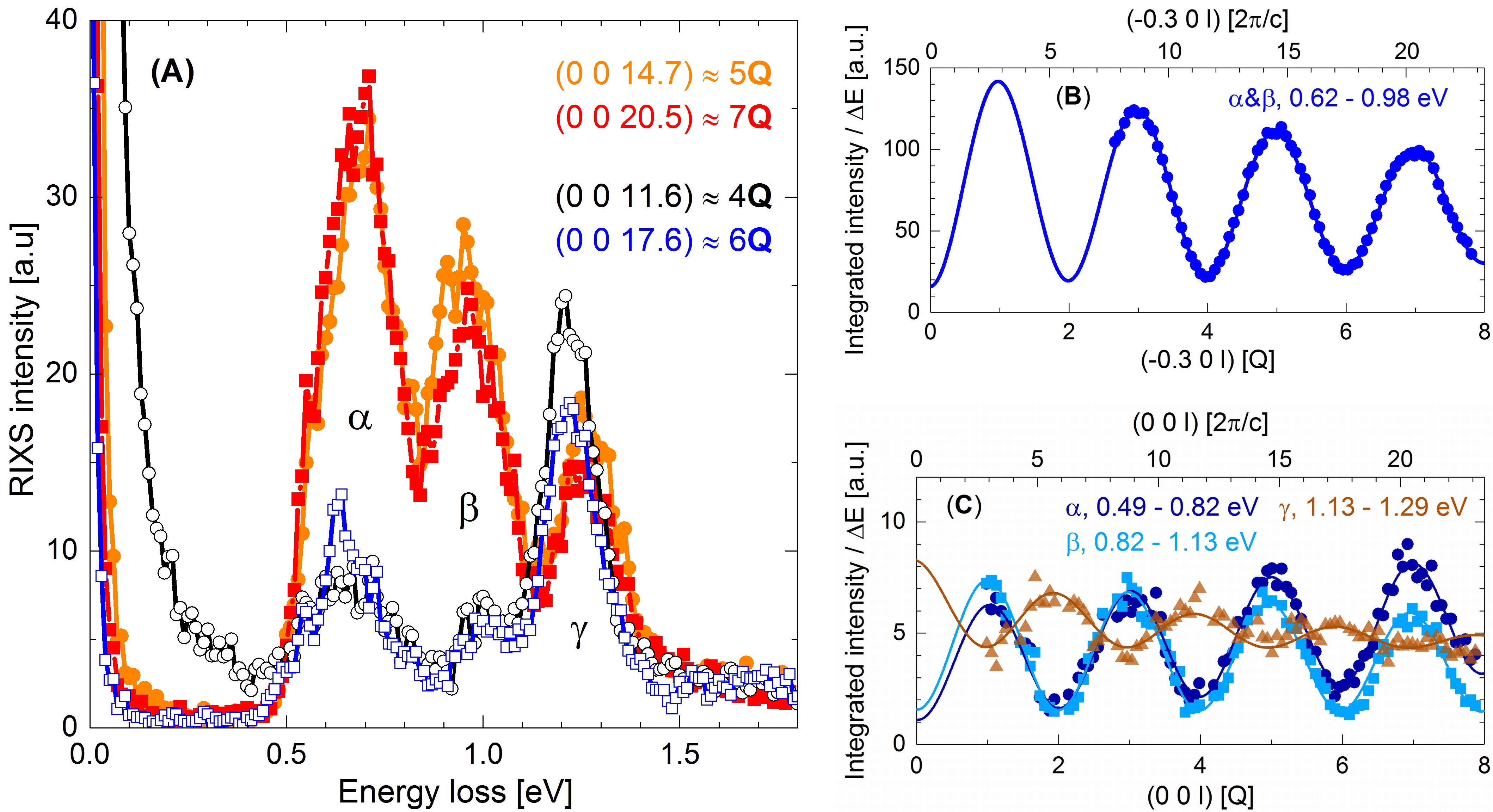}
	\caption{\textbf{RIXS data of intra-$t_{2g}$ excitations in BCIO.} 
		(\textbf{A}) High-resolution RIXS spectra  at $T$\,=\,20\,K for different transferred momenta 
		$\mathbf{q} =$(0 0 $mQ$) with integer $m$ and $Q$\,=\,$\pi/d$\,=\,$2.914 \cdot 2\pi/c$, 
		where $d$ and $c$ denote the intra-dimer Ir-Ir distance and the lattice constant, respectively. 
		The spectra show a pronounced even/odd behavior with respect to $m$, reflecting the sinusoidal {\bf q} dependence. 
		Note that the elastic peak is suppressed in $\pi$ polarization for a scattering angle of $2\theta$\,=\,$90^\circ$, 
		and that the data for $4Q$ to $7Q$ were measured at $2\theta$\,=\,$52^\circ$, $67^\circ$, $83^\circ$, 
		and $101^\circ$, respectively. Accordingly, the elastic peak is strongest for $4Q$.
		(\textbf{B,C})
		Interference patterns in the RIXS intensity as a function of $\mathbf{q}$. 
		The data cover about 3.5 periods in $Q$, equivalent to more than 20 Brillouin zones (top axis). 
		The intensity was integrated over the energy-loss ranges indicated in the figure and normalized 
		by the width $\Delta E$ of the respective energy range. 
		Data in (\textbf{B}) were measured at 10\,K with lower energy resolution of 0.36\,eV, 
		integrating over features $\alpha$ and $\beta$, to enhance the signal-to-noise ratio. 
		The high-resolution data in (\textbf{C}) discriminate between the three peaks $\alpha$ (dark blue), 
		$\beta$ (light blue), and $\gamma$ (brown). 
		Solid lines depict fits using $a_0 e^{-a_1 l}\cdot \sin^2(\pi l/2Q+\varphi) + b_0 + b_1 l$ with the 
		parameters $Q$, $a_0$, $a_1$, $b_0$, and $b_1$ as well as $\varphi$\,=\,0 or $\pi/2$. 	    
	}
	\label{fig_cos}
\end{figure}

Figure 3A depicts high-resolution RIXS spectra of BCIO for a fixed incident energy tuned to the 
Ir $L_3$ edge ($2p \! \rightarrow \! 5d$), see \textit{Methods}, 
which resonantly enhances inelastic scattering from intra-$t_{2g}$ excitations. 
With a $5d$ \mbox{$t_{2g}-e_g^\sigma$} splitting of about 3 eV,  
the observed features between 0.5 eV and 1.5 eV, labeled $\alpha$, $\beta$, and $\gamma$, 
can safely be attributed to intra-$t_{2g}$ excitations. 
The absence of dispersion strongly supports a local character, see appendix.
The spectra clearly do not display the characteristic feature of individual $j=1/2$ moments, 
the narrow spin-orbit exciton peaking at about $1.5 \lambda$. A textbook example of the latter 
is found in isostructural Ba$_3$Ti$_{2.7}$Ir$_{0.3}$O$_9$, 
where the small Ir content prevents dimer formation, see appendix.

The crucial observation is that the integrated intensity of the observed features shows the 
characteristic two-beam interference pattern \cite{Ma95,Gelmukhanov94}, 
i.e., a pronounced sinusoidal oscillation as a function of $q_c$, the component of the 
transferred momentum $\bf{q}$ parallel to the dimer axis, see Fig.\ 3B. 
The period $2Q=2\pi/d$ yields $d=2.530(8)$~\AA{} at $T=20$ K, in very good agreement 
with the Ir-Ir distance of $2.5361(7)$~\AA{} determined at $300$ K by x-ray diffraction, 
see appendix. 
Note that $Q=2.914 \cdot 2\pi/c$ is incommensurate with the reciprocal lattice vector $2\pi/c$, 
where $c$ denotes the lattice constant. 
A clear dichotomy with respect to even/odd $Q$ is also evident from the RIXS spectra in Fig.~3A.
Even without further knowledge of the underlying microscopic physics, the observed interference pattern 
with a period given by the Ir-Ir distance is an unmistakable proof of the double-slit-type RIXS process 
originating from the quasi-molecular orbital character of the investigated states.

\section{Discussion}

The most remarkable feature of RIXS  interferometry is the ability to determine the symmetry 
of the low-energy excitations  
and thus to distinguish between the two limiting scenarios sketched in Figs.~2C and 2D.
The symmetry is encoded in the phase of the interference pattern. 
Young's canonical elastic double-slit experiment gives a maximum for {\bf q}=0, 
which is equivalent to the $\cos^2(qd/2)$ dependence observed for peak $\gamma$, see Fig.~3C. 
The cosine denotes the Fourier transform of the double slit, i.e., its structure factor.
Strikingly, features $\alpha$ and $\beta$ show a $\sin^2(qd/2)$ modulation. 
In our experiment on structural dimers with a bonding singlet ground state, the $\sin^2(qd/2)$  
[$\cos^2(qd/2)$] behavior of the observed RIXS features reflects the bonding [antibonding] character of 
the corresponding excited-state wavefunctions. This behavior embodies the simple dipole selection rules 
for both the absorption and reemission processes, see appendix.  
The symmetry of the observed interference patterns agrees with Fig.~2D 
and suggests that this simplified sketch provides a valid starting point for an intuitive understanding 
of the three low-energy features $\alpha$, $\beta$, and $\gamma$.

This assignment is corroborated by careful modeling of the Ir$_2$O$_9$ bioctahedra. 
We take into account spin-orbit coupling $\lambda$, hopping terms within a dimer, 
the trigonal crystal field $\Delta_{\rm CF}$, and on-site Coulomb correlations described by Hubbard $U$ 
and Hund exchange $J_H$, see \textit{Methods}.
The calculated RIXS spectra and the corresponding $\mathbf{q}$-dependent intensities 
(see Fig.~4) qualitatively agree with our experimental results. 
In particular, we reproduce the $\sin^2(qd/2)$ behavior of peaks $\alpha$ and $\beta$  
and the $\cos^2(qd/2)$ behavior of peak $\gamma$. 
This implies quasi-molecular orbitals that are 
governed by a combination of strong spin-orbit coupling $\lambda \approx 0.4-0.5$ eV 
and strong hopping $\sim 1$ eV. 
The hopping is about 3 times larger than between nearest-neighbor $j=1/2$ moments in Sr$_2$IrO$_4$.
The dominant contribution to the ground state is a dimer singlet with $j_{\rm dim}=0$ 
in which the single-site $j=1/2$ moments occupy a bonding quasi-molecular orbital. 
As depicted in Fig.~2D, the two lower features $\alpha$ and $\beta$ correspond to excitations 
to bonding orbitals originating from the $j=3/2$ spin-orbit exciton, 
their splitting is mainly caused by $J_H$ and $\Delta_{\rm CF}$. 
We find that peak $\gamma$ at 1.2 - 1.3 eV corresponds to the antibonding triplet excitation from the 
$j_{\rm dim}=0$ ground state to a $j_{\rm dim}=1$ excited state of a dimer. 
This $j$ flip can be realized by both, spin-flip and spin-conserving processes. The former are allowed in 
$L$-edge RIXS due to the very strong spin-orbit coupling in the $2p$ shell in the intermediate state. 
In the calculated spectra the triplet excitation intensity is somewhat overestimated due to the neglect 
of the electron-hole continuum above the gap, which provides a possible decay channel 
for the local excitations considered here with a concomitant increase of the width.

\begin{figure}[t]
	\centering
	\includegraphics[width=0.7\columnwidth]{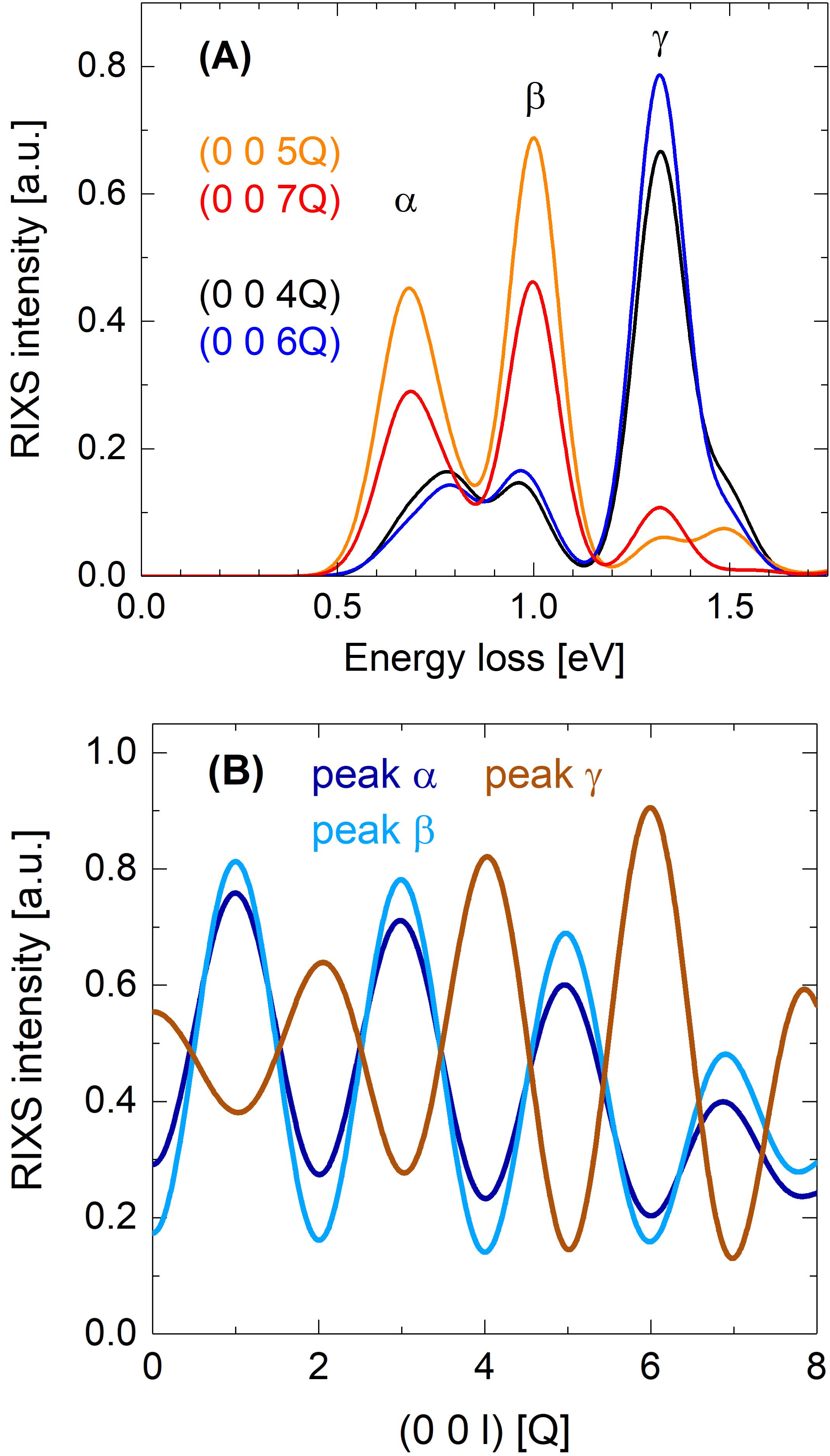}
	\caption{\textbf{Calculated RIXS data of BCIO.} 
		(\textbf{A}) Calculated RIXS spectra are plotted for the same $\mathbf{q}$ values as the experimental data in 
		Fig.~3, reproducing peaks $\alpha$, $\beta$, and $\gamma$ as well as their pronounced even/odd behavior with 
		respect to $Q$ (see appendix for more details). 
		(\textbf{B}) $\mathbf{q}$-dependent interference patterns for peaks $\alpha$, $\beta$, and $\gamma$, 
		which have to be compared with 
		the experimental result depicted in Fig.~3C.
		Parameters are $t_{a_{1g}}$\,=\,$1.1$\,eV, $t_{e_g^\pi}$\,=\,$0.5$\,eV, 
		$U$\,=\,$1.0$\,eV, $J_H$\,=\,0.3\,eV, $\Delta_{\rm CF}$\,=\,$-0.2$\,eV, and $\lambda$\,=\,$0.45$\,eV 
		(see appendix). 
	}
	\label{fig_CeSpecTheo}
\end{figure}

Finally we address the envelope of the sinusoidal interference pattern. 
The RIXS data cover about 20 Brillouin zones. Still the amplitude of the intensity modulation does not change 
strongly, at least for peaks $\alpha$ and $\beta$, allowing us to detect the interference pattern over a 
broad range of {\bf q}. 
This is due to the nearly point-like nature of the 'slits' formed by the core hole in the intermediate 
states \cite{Ament11}, i.e., the 'slit width' is negligible. 
The $\mathbf{q}$-dependent envelope rather contains valuable information on the dependence of the RIXS matrix elements 
on the scattering geometry and thereby on the precise wavefunctions. 

\section{Conclusion}

The comparison of theory and experiment on BCIO shows that the quasi-molecular orbitals, 
delocalized over the two Ir sites of a  dimer, 
give rise to the RIXS interference effect, which was predicted almost 25 years ago 
as an inelastic incarnation of Young's double-slit experiment.
Conceptually different interference effects as a function of energy but not of $q$ arise 
if \textit{energetically} different intermediate states -- located on the \textit{same} site -- 
contribute coherently \cite{Wray15,Pietzsch11}. 
In contrast, our experiment realizes the genuine spatial double-slit setup: 
a photon scatters inelastically on one of two dimer sites. 
Previously, this was discussed in the context of dimers in VO$_2$ \cite{He16}, 
but only a single value of $q$ was analyzed due to the limited momentum transfer available in the soft x-ray range. 
Similarly, studies of magnons in the bilayer compound Sr$_3$Ir$_2$O$_7$ \cite{MorettiSr3} and of stripes 
in nickelates and cuprates \cite{Schuelke11} addressed only a few values of $q$. 
Coverage of a broad range of $q$ fully reveals the interference character
and allowed us to unravel the symmetry and character of electronic dimer excitations in BCIO. 
These results demonstrate the potential of this interference method 
to probe the electronic structure of materials containing well-defined structural units such as dimers, 
trimers, or heptamers \cite{Streltsov17} 
as well as structures in which the carriers are 'localized' 
only in one direction, e.g., bilayers or ladders. 
More specifically, our results suggest that RIXS interferometry is ideally suited to explore the role 
of molecular orbitals in the spin-liquid candidate Ba$_3$InIr$_2$O$_9$ with In$^{3+}$ ions and 
\textit{three} holes per dimer \cite{Dey17} 
as well as to search for Majorana fermions in iridate candidates for a Kitaev spin liquid \cite{RauReview}. 
The latter quantum state lives on tricoordinated lattices, still spin correlations are restricted 
to nearest neighbors. 
Majorana fermion excitations are thus expected to show an interference pattern which is closely 
related to the case of dimers in BCIO.
In general, RIXS  interferometry will prove to be an efficient tool for studying the symmetry 
and character of excited stated in complex materials.

\section{Materials and Methods}

\textbf{Crystal growth and characterization.} 
The Ba$_3$$M$Ir$_2$O$_9$ family with the hexagonal (6H)-BaTiO$_3$ structure is formed for a wide variety 
of di-, tri-, or tetravalent $M$ ions \cite{Doi04,Sakamoto06,Dey12,Kumar16}. 
The valence states of $M$\,=\,Ce$^{4+}$ and Ti$^{4+}$ yield $5d^5$ Ir$^{4+}$ ions. 
Single crystals of Ba$_3$CeIr$_2$O$_9$ and Ba$_3$Ti$_{2.7}$Ir$_{0.3}$O$_9$
have been grown by the melt-solution technique and spontaneous nucleation. Using stoichiometric amounts
of BaCO$_3$, CeO$_2$, and IrO$_2$ for the growth of Ba$_3$CeIr$_2$O$_9$ and BaCO$_3$, TiO$_2$, and IrO$_2$ 
in a ratio of 3\,:\,2\,:\,1 in the case of Ba$_3$Ti$_{2.7}$Ir$_{0.3}$O$_9$, and in both cases an addition 
of BaCl$_2$\,$\cdot$\,2H$_2$O, hexagonal prismatic crystals of about $2$\,mm size were grown within 
growth periods of four weeks.
The crystals were mechanically separated from the flux, washed with cold H$_2$O,
and characterized by x-ray diffraction, energy-dispersive x-ray (EDX) analysis, 
and magnetization measurements. 

\textbf{RIXS measurements.} RIXS measurements at the Ir $L_3$ edge were performed on the ID20 beamline 
at ESRF \cite{Moretti18}.
We used an incident energy of 11.2155\,keV for $M$\,=\,Ce and 11.2150\,keV for $M$\,=\,Ti
to maximize the resonantly enhanced RIXS intensity of the intra-$t_{2g}$ excitations.
An overall energy resolution of 27\,meV was obtained by combining a Si(844) 
back-scattering monochromator and $R$\,=\,2\,m Si(844) spherical diced crystal analyzers \cite{Moretti13}. 
These analyzers make the RIXS setup partially dispersive, i.e., 
one image on the pixelated detector covers an energy range of approximately 360\,meV. 
Accordingly, the RIXS spectra of Fig.~3A were measured by scanning the energy at constant {\bf q} 
while the data in Fig.~3C were collected by scanning {\bf q} at constant energy.
The data in Fig.~3B were measured with a lower resolution of 0.36\,eV using a Si(311) channel-cut 
in place of the Si(844) back-scattering monochromator in order to enhance the signal-to-noise ratio.
For both $M$\,=\,Ce and Ti, RIXS measurements were performed on polished (0 0 1) surfaces 
with the $c$ axis in the horizontal scattering plane and the $a$ axis along the vertical direction. 
The incident photons were $\pi$ polarized.
Samples were cooled using a continuous-flow He cryostat.

\textbf{Theory.} 
The sketch in Fig.~2D provides a simplified but intuitive picture of the low-energy excitations.
For the simulation of the RIXS spectra, we calculate all 66 eigenstates for two $t_{2g}$ holes 
on two Ir sites of an Ir$_2$O$_9$ bioctahedron via exact diagonalization. 
We start from the Hamiltonian for a single Ir site, taking into account spin-orbit coupling 
$\lambda \, \mathbf{S}\cdot \mathbf{L}$ and the trigonal crystal field,  $\Delta_{\rm CF} L_z^2$. 
The on-site Coulomb repulsion between carriers with different spins 
in the same or in different orbitals is given by $U$ and $U-2J_H$, respectively, and 
the interaction Hamiltonian reads 
\begin{eqnarray}
\nonumber
H_C & = & U \, \sum_{i,\alpha} n_{i\alpha \uparrow} n_{i\alpha \downarrow} 
+  \frac{1}{2} (U-3J_H) \sum_{i,\sigma,\alpha\neq\alpha^\prime} n_{i\alpha \sigma} n_{i\alpha^\prime \sigma} \\
\nonumber
& + & (U-2J_H) \sum_{i,\alpha\neq\alpha^\prime} n_{i\alpha \uparrow} n_{i\alpha^\prime \downarrow} 
\\
\nonumber
& + & (U-2J_H)\sum_{i} \Big( 15-5\sum_{\alpha,\sigma} n_{i\alpha \sigma} \Big) \, ,
\end{eqnarray}
where $n_{i\alpha\sigma}$ is the number operator for the $t_{2g}$ orbitals 
$\alpha \in \{a_{1g},e_g^{\pi+},e_g^{\pi-}\}$ \cite{KhomskiiZhETF} at site $i$ with $\sigma \in \{\uparrow,\downarrow\}$. 
Additionally, we consider the hopping interactions $t_{a_{1g}}$ and $t_{e_g^\pi}$ between two Ir sites.

The two octahedra forming an Ir$_2$O$_9$ bioctahedron are rotated by 180$^\circ$ with respect 
to each other around the trigonal $z$ axis, see Fig.\ 2(A) and Fig.\ S1 in the appendix. 
This causes a sign change of the $xz$ and $yz$ orbitals in one of the two octahedra, which affects 
the selection rules of excitations involving the $e_g^{\pi}$  orbitals \cite{KhomskiiZhETF}.
For the calculation of the RIXS spectra, we consider the RIXS matrix elements for single-particle excitations 
between all 66 states in the dipole approximation for both absorption and reemission. 
As typical for $t_{2g}^5$ iridates \cite{Ament11b}, we neglect multiplet effects between 
the intermediate-state $2p_{3/2}$ core hole and $5d$ holes, which significantly simplifies the calculation. 
It is motivated for a $t_{2g}^5$ configuration by the full $t_{2g}$ shell in the intermediate state, 
neglecting the $e_g^\sigma$ holes. Note, however, that spin-orbit coupling mixes $e_g^\pi$ and $e_g^\sigma$ levels. 
Moreover, this approximation is not fully valid for quasi-molecular orbitals delocalized over two Ir sites, 
where states with two holes on the same site contribute. 
For comparison with experiment, one has to add the RIXS intensities of the two layers 
of bioctahedra present in Ba$_3$CeIr$_2$O$_9$, see Fig.\ 2(A), which for the present geometry is equivalent 
to adding intensities for $+q$ and $-q$. Finally, RIXS spectra are calculated for certain $q$ values by 
assuming Gaussian line shape with a fixed width.

\textbf{Acknowledgments.} 
We gratefully acknowledge fruitful discussions with M. Braden, I. Mazin, and S. Trebst. 
This project was funded by the Deutsche Forschungsgemeinschaft (DFG, German Research Foundation) 
-- Project numbers 277146847 and 247310070 -- CRC 1238 (projects A02, B01, B02, B03, and C02) 
and CRC 1143 (project A05), respectively. 
The work of S.V.S. was supported by the Russian Science Foundation through the project 17-12-01207.

\renewcommand{\theequation}{S\arabic{equation}}
\renewcommand{\thefigure}{S\arabic{figure}}
\renewcommand{\thetable}{S\,\Roman{table}}

\setcounter{figure}{0}

\appendix

\section{Appendix A: Crystal structure}

\begin{figure}[tb]
	\centering
	\includegraphics[width=35mm]{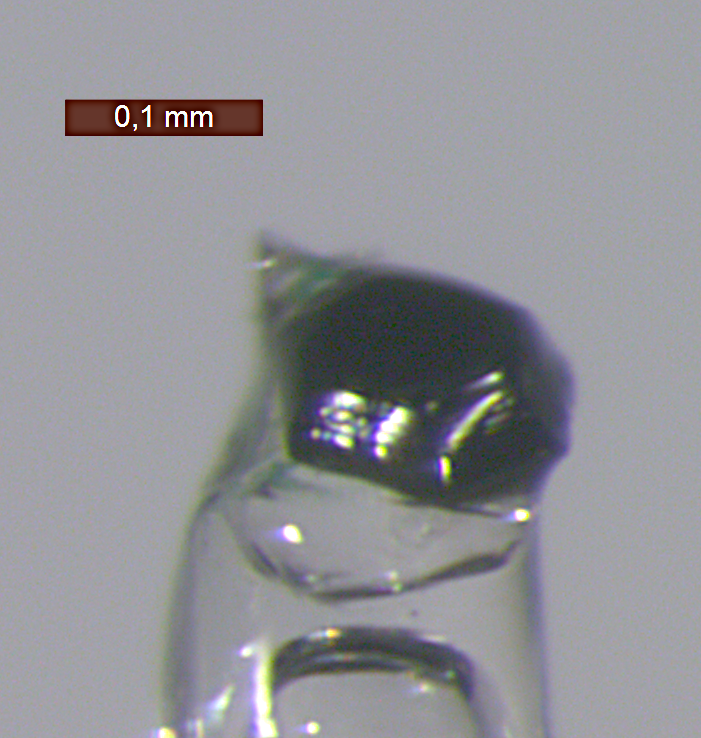}
	\hspace{7mm}
	\includegraphics[width=38mm]{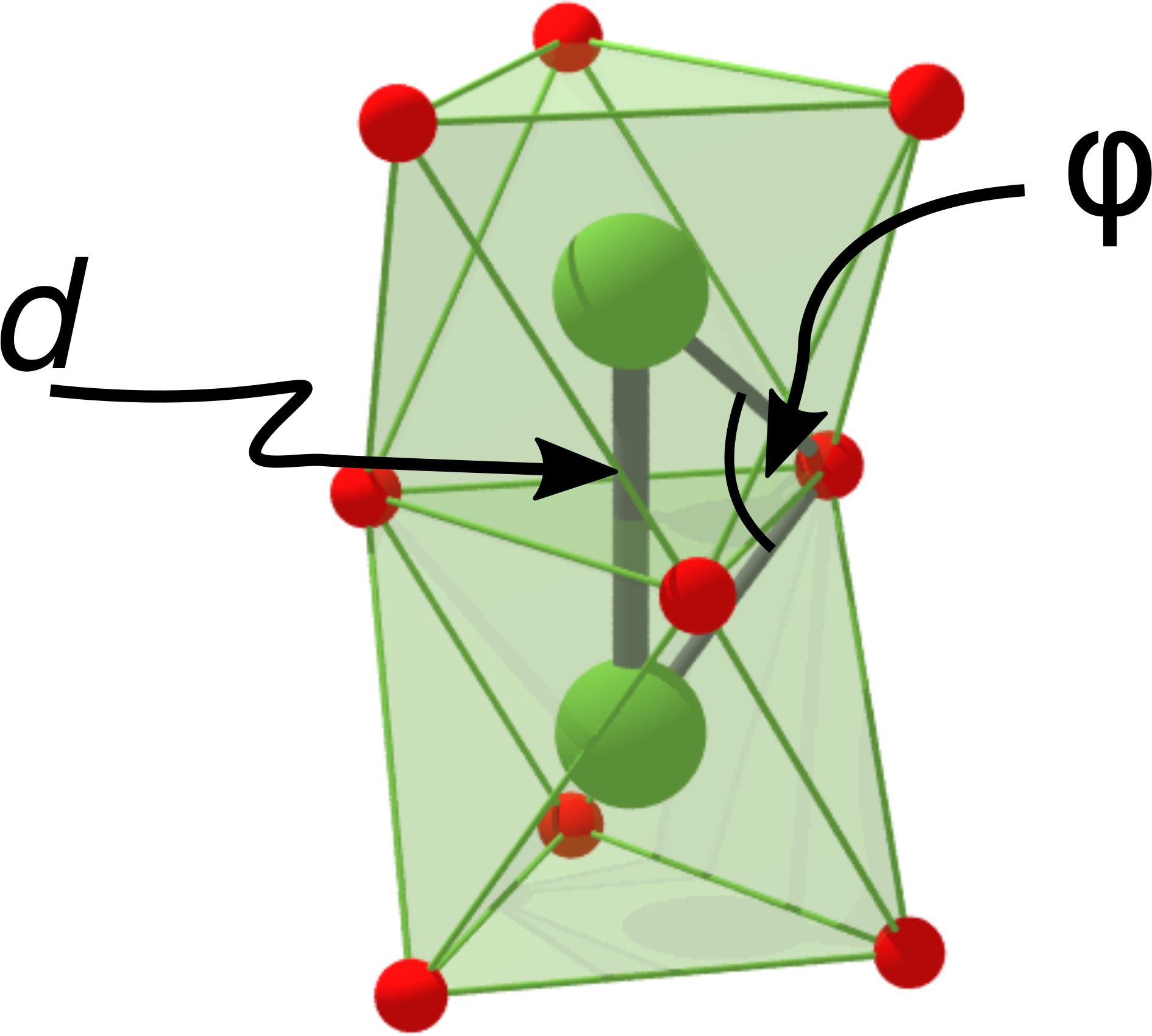}
	\caption{\textbf{Sample of \ceir{} used for single-crystal x-ray diffraction (left) 
			and sketch of an Ir$_2$O$_9$ bioctahedron with face-sharing octahedra (right).} 
		At 300\,K, the Ir-Ir bond distance equals $d$\,=\,$2.5361(7)$\,\AA{} 
		and the Ir-O-Ir angle is $\varphi$\,=\,$78.285(2)^\circ$. }
	\label{fig_sample}
\end{figure}

The crystal structure of Ba$_3$CeIr$_2$O$_9$ was analyzed by x-ray diffraction measurements. 
Small crystals were ground to a fine powder which was studied on a D5000 diffractometer (Siemens) 
with Cu $K_{\alpha}$ radiation at 300\,K.\@ 
In order to calibrate the lattice constants, a small amount of Si powder was mixed with the sample. 
The lattice parameters were determined by Rietveld refinements with the software FullProf \cite{Rodriguez1993a}. 
We find $a$\,=\,$b$\,=\,$5.9035(9)$\,{\AA} and $c$\,=\,$14.715(3)$\,{\AA} in good agreement with the 
powder diffraction analysis reported by Doi and Hinatsu \cite{Doi04}.

\begin{table}[b]
	\begin{tabular}{c | c | c | c | r}
		& $ x $   &  $ y $       &    $ z $    & $ U_\mathrm{iso} $ or $ U_\mathrm{ani} $   \\\hline
		Ba(1) &  0         &  0         &  1/4         &  0.00822(12) \\
		Ba(2) &  1/3  &  2/3  &  0.09993(4)   &  0.00970(11) \\
		Ir    &  1/3  &  2/3  &  0.663825(17) &  0.00400(7)  \\
		Ce    &  0         &  0         &  1/2          &  0.00388(11) \\
		O(1)  &  0.1742(3) &  0.8258(3) &  0.5870(3)    &  0.0107(5)   \\
		O(2)  &  0.4857(4) &  0.5143(4) &  3/4         &  0.0052(5)
	\end{tabular}\\
	\vspace{5mm}
	\begin{tabular}{c | c | c | c | c }
		&    $ U_{11} $  &  $ U_{22} $  &  $ U_{33} $ &    $ U_{23} $ \\\hline
		Ba(1) &  0.00917(15) &  0.00917(15) &  0.0063(2) &  0.00458(7)     \\
		Ba(2) &  0.00793(12) &  0.00793(12) &  0.0132(2) &  0.00397(6)     \\
	\end{tabular}
	\caption{\textbf{Structural parameters determined by x-ray diffractometry.}
		Positions are given in units of the cell parameters. The atomic isotropic or anisotropic 
		displacement parameters $U_\mathrm{iso}$, $U_\mathrm{ani}$, and $U_{ij}$ are given in {\AA}$^2$. 
		$U_{12}$\,=\,$U_{13}$\,=\,$0$\,{\AA}$^2$ are constrained by symmetry.}
	\label{tab_xray_structure}
\end{table}

The single-crystal diffraction experiment was performed on a Bruker AXS Kappa APEX II four-circle 
x-ray diffractometer with a wavelength of $0.71073$\,{\AA} (Mo K$_{\alpha}$). 
A non-spherical sample with a spatial extension of roughly $100$\,$\mu$m (see Fig.\ \ref{fig_sample}) 
was investigated at room temperature. Using a distance of $40$\,mm between detector and sample, 
we collected $94\,002$ reflections, of which $1\,395$ are unique. $1\,241$ of these reflections have 
an intensity significantly greater than zero. The absorption correction was carried out using a 
non-spherical model containing $25$ faces and an absorption coefficient of $\mu$\,=\,$48$\,mm$^{-1}$, 
yielding a weighted internal R value of $wR2(\mathrm{int})$\,=\,$5.8$\%. 
The structural model according to space group $P 6_3/m m c$ (Nr.\ $194$) with $17$ structural parameters 
was refined using the software Jana2006 \cite{Petricek}. 
The refinement further includes an isotropic extinction correction.
The structural refinement yields $R(F)$ values of $R(\mathrm{obs})$\,=\,$4.37$\%, $wR(\mathrm{obs})$\,=\,$6.06$\%, 
$R(\mathrm{all})$\,=\,$5.01$\%, and $wR(\mathrm{all})$\,=\,$6.14$\%. 
In the refinements we allowed for a shared occupation of the Ce and Ir sites with the constraint of 
full total occupation.
However, the deviation of the refined occupation numbers from the ideal ones are small and negative, 
namely $-0.029(12)$ for Ce on an Ir site and $-0.019(10)$ for Ir on a Ce site. There is thus no evidence 
for any disorder between Ce and Ir occupation. 
Only the atomic displacement parameters of the Ba  atoms are refined anisotropically. 
The atomic positions and displacement parameters are listed in Table \ref{tab_xray_structure}.
The Ir-Ir bond distance amounts to $d$\,=\,$2.5361(7)$\,{\AA} and the Ir-O-Ir angle to 
$\varphi$\,=\,$78.285(2)^\circ$ (see Fig.\ \ref{fig_sample}), 
larger than the value of $\varphi_0 \approx 70.5^\circ$ expected for undistorted octahedra. 
Doi and Hinatsu \cite{Doi04} 
reported a similar value of $d$\,=\,$2.5266(17)$\,{\AA}, also at room temperature.

\section{Appendix B: RIXS on B\MakeLowercase{a}$_3$T\MakeLowercase{i}$_{2.7}$I\MakeLowercase{r}$_{0.3}$O$_9$}

\begin{figure}[b]
	\centering
	\includegraphics[width=.9\columnwidth]{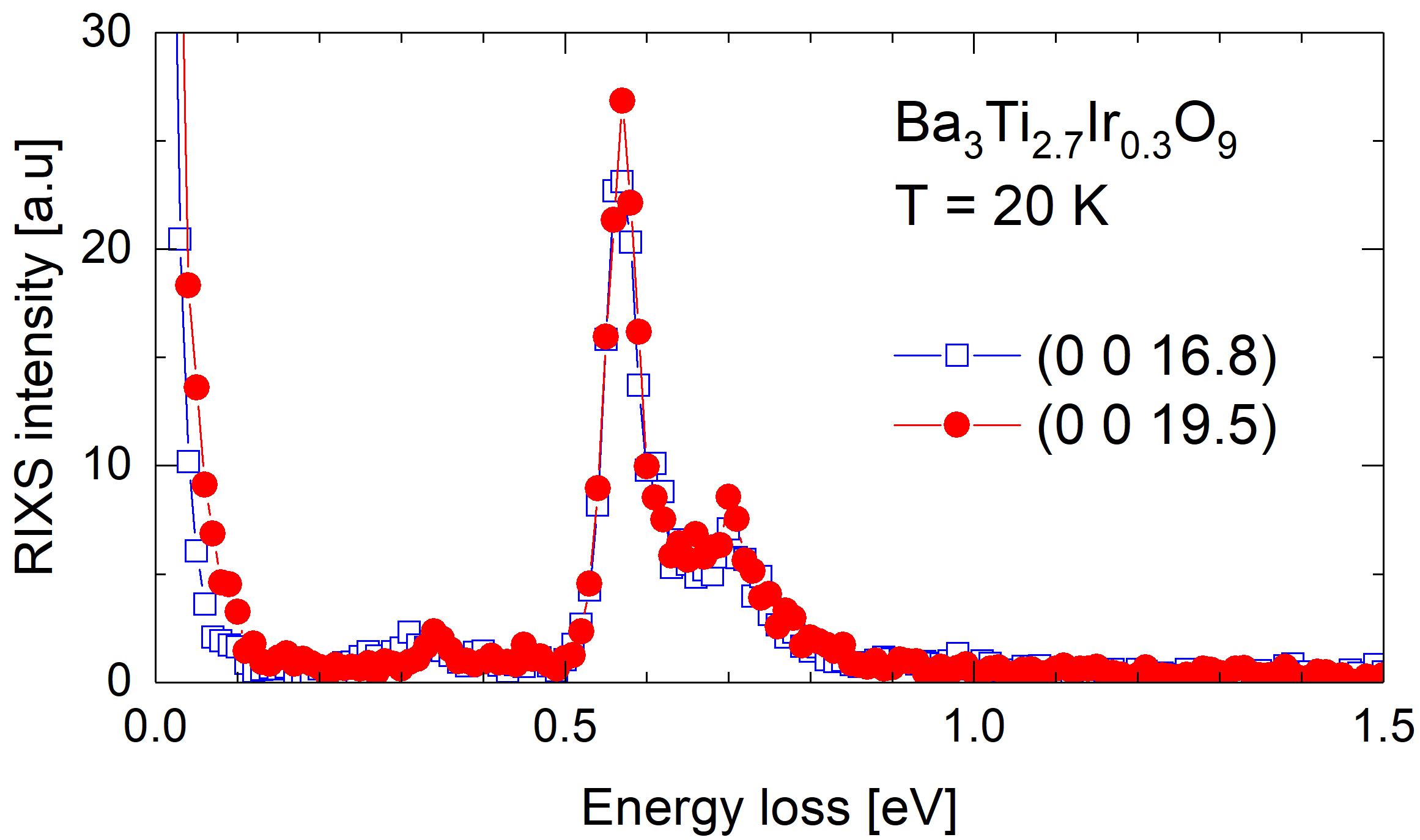}
	\caption{\textbf{RIXS spectra of Ba$_3$Ti$_{2.7}$Ir$_{0.3}$O$_9$.} 
		The small Ir content yields individual Ir$^{4+}$ sites instead of dimers, showing a textbook example 
		of the spin-orbit exciton, very different from 	the dimer spectra of Ba$_3$CeIr$_2$O$_9$. }
	\label{fig_RIXS_Ti}
\end{figure}

The RIXS spectra of \tiirthree{} show two narrow peaks at 0.57\,eV and 0.70\,eV, see Fig.\ \ref{fig_RIXS_Ti}. 
Considering the narrow line width, the comparably small splitting of 0.13\,eV, and the insensitivity to the 
transferred momentum $\mathbf{q}$, the data of \tiirthree{} are a textbook example of the spin-orbit exciton 
of individual $j$\,=\,$1/2$ moments (in a non-cubic field), very similar to the case of iridium fluorides with 
strongly localized states \cite{Rossi17} 
and very different from the dimer spectra observed in Ba$_3$CeIr$_2$O$_9$. 
This results from the small Ir content $x$\,=\,$0.3$ with only 15\,\% of the bioctahedral metal sites 
occupied by Ir$^{4+}$ ions. The small value of $x$ suppresses the formation of Ir$_2$O$_9$ units, thus 
nearest-neighbor Ir$^{4+}$ ions are isolated in IrTiO$_9$ bioctahedra. 
Based on a local crystal-field calculation \cite{Sizyuk14}, 
the two peak energies yield two possible solutions, either 
$\lambda$\,=\,$0.40$\,eV and a trigonal crystal field of $\Delta_{\rm trig}$\,=\,-0.23\,eV 
or $\lambda$\,=\,$0.41$\,eV and $\Delta_{\rm trig}$\,=\,+0.18\,eV.\@
In the convention of our Hamiltonian, the negative sign of $\Delta_{\rm trig}$\,=\,-0.23\,eV  
agrees with the elongated octahedra found in x-ray diffraction. 
In hole notation, it corresponds to a lowering of the $a_{1g}$ orbital, as depicted in Fig.\ 2.

\section{Appendix C: Absence of significant dispersion}

The scenario of local dimers is well supported by the absence of any significant dispersion 
upon variation of the transferred momentum parallel or perpendicular to $c$, see Fig.~S3. 
The strong change of intensity upon variation of {\bf q} parallel to $c$ (bottom panel) agrees with the data 
shown in Fig.~3 of the main text, it reflects double-slit interference. 
In contrast, the small monotonic change of intensity of feature $\beta$ at 1 eV observed for changing {\bf q} 
perpendicular to $c$ (top panel) can be attributed to matrix element effects, i.e., 
to polarization and experimental geometry.

\begin{figure}[h]
	\centering
	\includegraphics[width=0.8\columnwidth]{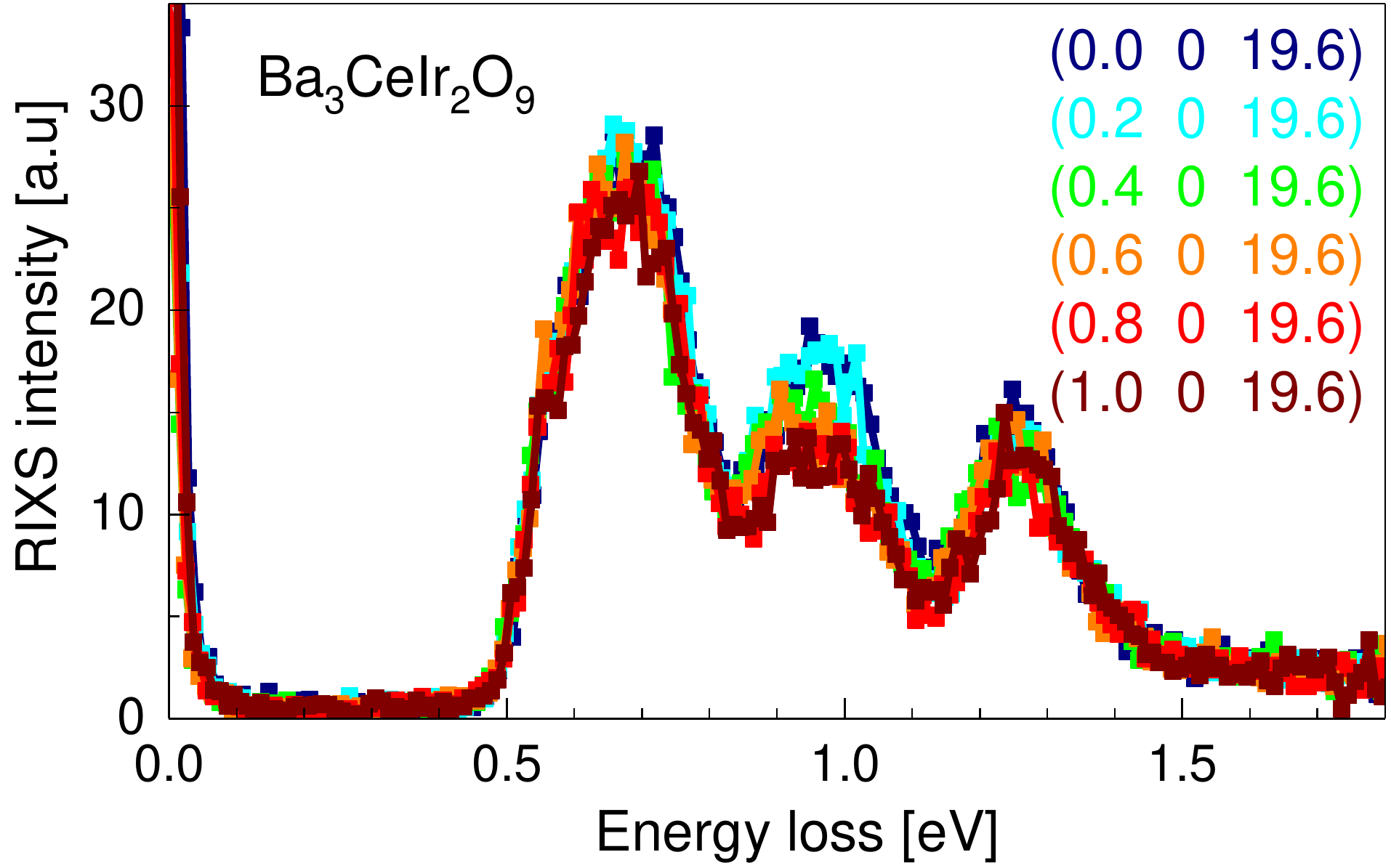}
	\includegraphics[width=0.8\columnwidth]{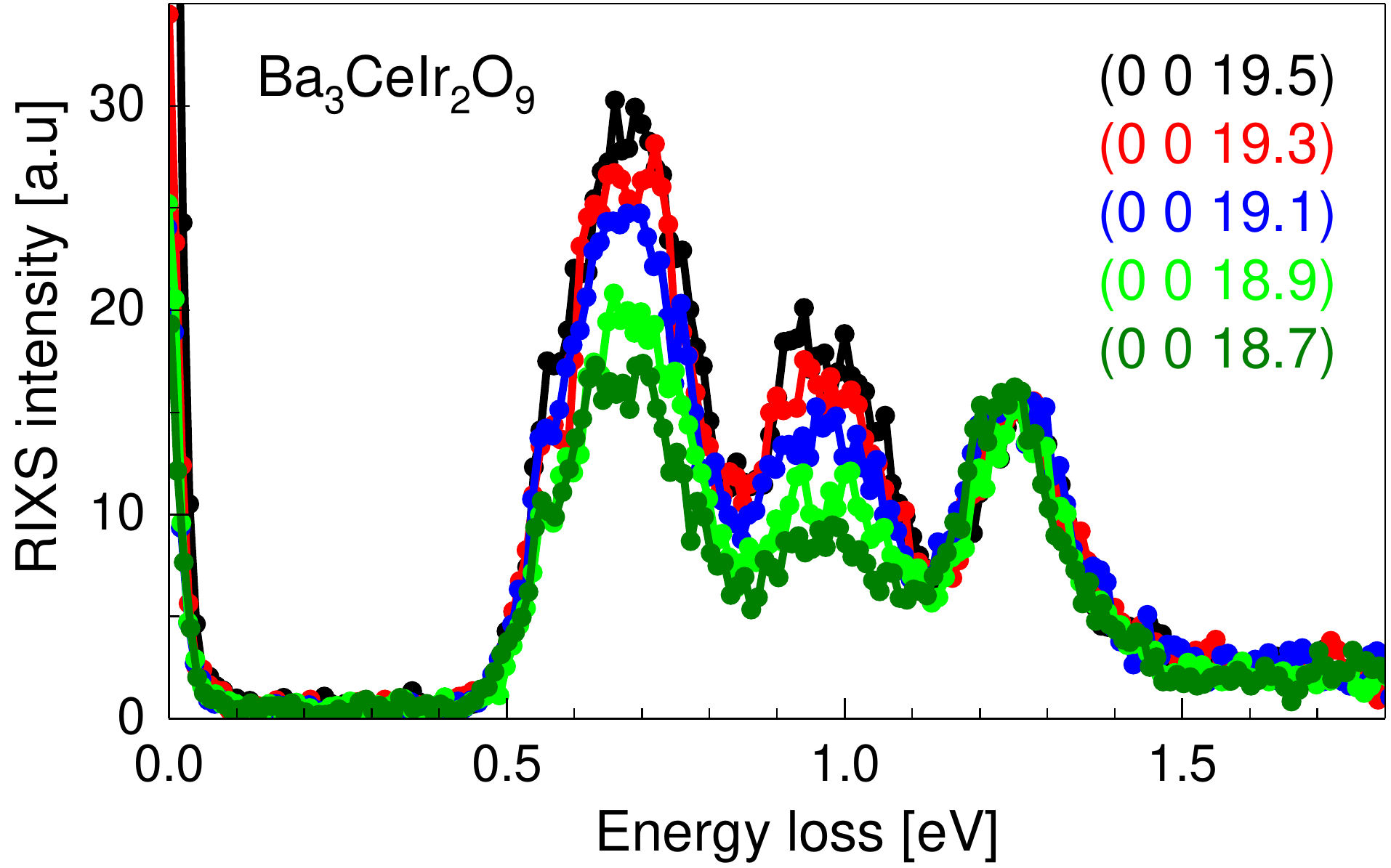}
	\caption{\textbf{Absence of dispersion in the RIXS spectra.} 
		The three RIXS features do not show any significant dispersion upon variation of 
		the transferred momentum either parallel (bottom) or perpendicular (top) to $c$. }  
	\label{fig_dispersion}
\end{figure}

\section{Appendix D: Momentum-dependence of double-slit-type RIXS}

The sinusoidal $\mathbf{q}$ dependence is obtained in a straightforward way by considering the RIXS intensity 
$I(\mathbf{q},\omega)$\,=\,$\sum_f |A_f(\mathbf{q})|^2 \, \delta(\hbar\omega - E_f)$ 
for the excited states $|f\rangle$ with energy $E_f$.
In dipole approximation, the amplitude $A_f(\mathbf{q})$ reads \cite{Ament11b}
\begin{equation}
\nonumber
A_f(\mathbf{q})\propto \langle f |\, \sum_\mathbf{R} e^{i\mathbf{q}\mathbf{R}}\, 
\left[ D^\dagger (\varepsilon^*_{\rm out})D(\varepsilon_{\rm in}) \right]_R \,| 0\rangle
\end{equation}
where $|0\rangle$ denotes the ground state, $\varepsilon_i$ the polarization of incident and scattered photons, 
$D$ the local dipole transition operator, and $\mathbf{R}$ runs over all Ir sites that contribute to a given final state $|f\rangle$. 
In the present case with quasi-molecular orbitals within a bioctahedron, these are the two 
crystallographically equivalent Ir sites at $\mathbf{r}_{1,2} = (0, 0, \pm d/2)$. 
For these two sites, the matrix elements may only differ in sign. 
For $\mathbf{q} \parallel \mathbf{r}_{2}-\mathbf{r}_1$, the even and odd combinations 
$[D^\dagger D]\,(e^{i\mathbf{q}\mathbf{r}_1} \pm e^{i\mathbf{q}\mathbf{r}_2})$ yield \cite{Ma95}
\begin{eqnarray}
\nonumber
A_f^+(q) & \propto & \cos(qd/2) \, \langle f |\, [D^\dagger (\varepsilon^*_{\rm out})D(\varepsilon_{\rm in})] \,| 0\rangle \, ,
\\ \nonumber
A_f^-(q) & \propto & \sin(qd/2)  \, \langle f |\, [D^\dagger (\varepsilon^*_{\rm out})D(\varepsilon_{\rm in})] \,| 0\rangle   \, , 
\end{eqnarray}
which explains the observed $\mathbf{q}$ dependence, in particular the period $2Q$\,=\,$2\pi/d$ in the intensity.

A precise determination of electronic parameters is beyond the scope of this study, 
but we consider the parameters used in the main text as very reasonable. 
Our values for the intra-orbital $U$\,=\,$1.0$\,eV, $J_H$\,=\,0.3\,eV, $\lambda$\,=\,$0.45$\,eV, and 
$\Delta_{\rm CF}$\,=\,$-0.2$\,eV lie well within the accepted range for iridates \cite{KimKhalNa14}. 
In the convention of our Hamiltonian, the negative sign of $\Delta_{\rm CF}$ 
corresponds to the situation depicted in Fig.\ 2, where the $a_{1g}$ orbital is lowered 
in the hole picture. 
For an estimate of the hopping parameters, we performed band-structure calculations 
in the generalized gradient approximation (GGA) \cite{Perdew96}. 
These calculations yield 
$t_{a_{1g}}$\,$\approx$\,$0.9$\,eV and $t_{e_g^\pi}$\,$\approx$\,$0.3$\,eV.\@ We used 
$t_{a_{1g}}$\,=\,$1.1$\,eV and $t_{e_g^\pi}$\,=\,$0.5$\,eV for the simulation of the RIXS data, 
in reasonable agreement with the prediction. 
Small deviations may partially be caused by neglecting the empty $e_g^\sigma$ orbitals, 
which is the typical approach for $5d^5$ iridates. Both, spin-orbit coupling and 
Coulomb correlations give rise to a finite mixing of $e_g^\pi$ and $e_g^\sigma$ orbitals, affecting the effective 
hopping parameter. Furthermore, our simplified two-site model neglects oxygen states and the electron-hole 
continuum, impeding a full quantitative description of the RIXS spectra.

\section{Appendix E: Substructure of the three RIXS features}

Figure 2D of the main text is a simplified sketch which, however, provides an appropriate description 
of the essential physics for the low-energy excitations discussed here.
As an example, consider the following ground-state wavefunction obtained from 
the comparison of experiment and theory, i.e., using the parameters given in Fig.~4 of the main text, 
\begin{eqnarray}
\nonumber
\Psi_0 & \approx & \alpha \, |\frac{1}{2},\frac{1}{2}\rangle\, |\frac{1}{2},-\frac{1}{2}\rangle 
\\ \nonumber
& + & \beta \, \left[ e^{i\frac{\pi}{2}}|\frac{1}{2},\frac{1}{2}\rangle\, |\frac{3}{2},\frac{3}{2}\rangle  + e^{-i\frac{\pi}{2}}|\frac{1}{2},-\frac{1}{2}\rangle\,|\frac{3}{2},-\frac{3}{2}\rangle  \right] 	
\\ \nonumber
& + & \gamma \, \left[e^{i\frac{\pi}{4}} |\frac{1}{2},\frac{1}{2}\rangle\, |\frac{3}{2},\frac{1}{2}\rangle  - e^{-i\frac{\pi}{4}}|\frac{1}{2},-\frac{1}{2}\rangle\,|\frac{3}{2},-\frac{1}{2}\rangle  \right] 
\\ \nonumber
& + & \delta \, \left[ |\frac{1}{2},\frac{1}{2}\rangle\, |\frac{3}{2},-\frac{1}{2}\rangle  + |\frac{1}{2},-\frac{1}{2}\rangle\,|\frac{3}{2},\frac{1}{2}\rangle  \right] 	
\\ \nonumber
& + & \epsilon \, \left[e^{-i\frac{\pi}{4}} |\frac{1}{2},\frac{1}{2}\rangle\, |\frac{3}{2},-\frac{3}{2}\rangle  - e^{i\frac{\pi}{4}}|\frac{1}{2},-\frac{1}{2}\rangle\,|\frac{3}{2},\frac{3}{2}\rangle  \right] \, .	
\end{eqnarray}
All terms refer to bonding (i.e., even) combinations of $|j,j_z\rangle\,|j^\prime,j^{\prime}_z\rangle$ with  
$\alpha$\,=\,0.82, $\beta$\,=\,$0.23$, $\gamma$\,=\,-0.27, $\delta$\,=\,0.02, and $\epsilon$\,=\,0.17. 
Terms with amplitude $< 0.01$ are omitted for clarity. 
The strongly dominant value of $\alpha$ supports the picture sketched in Fig.~2D of the main text.

The three features in the experimental RIXS spectrum are composed of several related excitations 
which are slightly split by the interplay of Coulomb interactions, crystal field, and spin-orbit coupling. 
Figure S4 resolves the individual excitations, using the same parameters as in the main text and a reduced 
Gaussian width. The experimentally unresolved substructure gives rise to an asymmetric line shape and 
causes small changes of shape and energy of the experimental features as a function of $q$. 
Overall, the character of the experimentally resolved excitations is well approximated by Fig. 2D. 

\begin{figure}[b]
	\centering
	\includegraphics[width=.8\columnwidth]{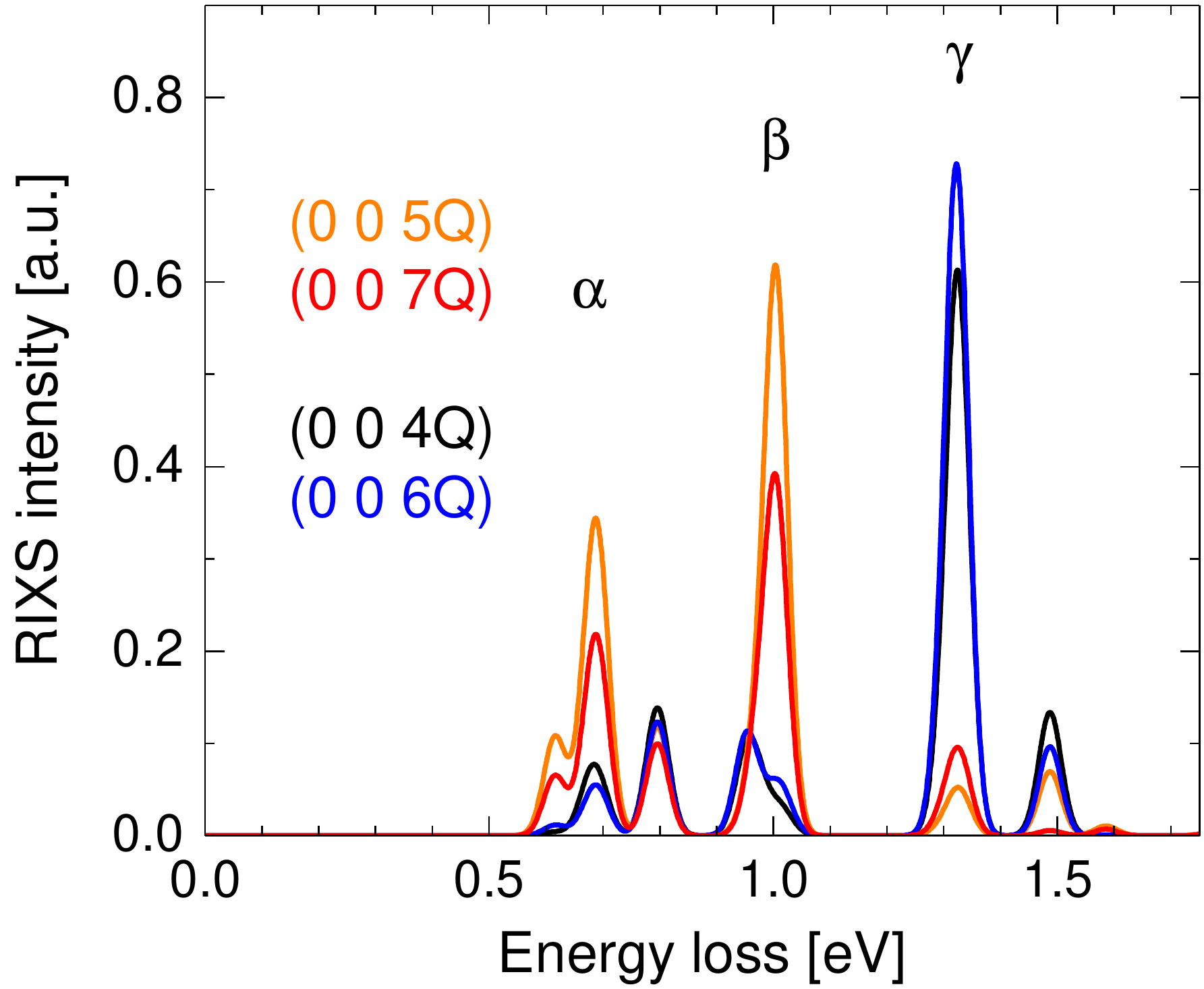}
	\caption{\textbf{RIXS spectra calculated with reduced width.} 
		Parameters are the same as in Fig.~4 of the main text, only the Gaussian width has been reduced.}  
	\label{fig_theo_spectra_highres}
\end{figure}

\end{document}